\documentclass[jnr,crcready,onecolumn]{iosart2x}
\makeatletter
\let\numberlines@hook\relax
\makeatother
\usepackage{graphicx}
\usepackage{float}
\usepackage{color}
\usepackage{amssymb}

\pubyear{0000}
\volume{0}
\firstpage{1}
\lastpage{1}

\begin{document}

\begin{frontmatter}

\title{Development of neutron scattering kernels for cold neutron reflector materials}
\runtitle{Development of neutron scattering kernels for cold neutron reflector materials}


\author[A]{\inits{J.R.}\fnms{Jos\'e Rolando}
  \snm{Granada}\ead[label=e1]{granada@cnea.gov.ar}%
  \thanks{Corresponding author. \printead{e1}.}},
\author[B]{\inits{J.I}\fnms{Jos\'e Ignacio} \snm{M\'arquez Dami\'an}},
\author[A,C]{\inits{J.}\fnms{Javier} \snm{Dawidowski}},
\author[C]{\inits{J.}\fnms{Jos\'e Ignacio} \snm{Robledo}}, 
\author[A]{\inits{C.H}\fnms{Christian} \snm{Helman}}, 
\author[D]{\inits{G.}\fnms{Giovanni} \snm{Romanelli}}, and
\author[D]{\inits{G.}\fnms{Goran} \snm{\v{S}koro}}
\address[A]{Departamento F\'{\i}sica de Neutrones, Centro At\'omico
  Bariloche, \orgname{Comisi\'on Nacional de Energ\'{\i}a At\'omica},
  RN, \cny{Argentina}\printead[presep={\\}]{e1}}
\address[B]{Spallation Physics Group, \orgname{European Spallation
    Source ERIC}, Lund, \cny{Sweden}}
\address[C]{\orgname{CONICET},
  \cny{Argentina}}
\address[D]{ISIS Facility, \orgname{Rutherford Appleton Laboratory},
  Chilton, Didcot, Oxfordshire OX11 0QX,\cny{U. K.}}

\begin{abstract}
  
  The newest neutron scattering applications are highly
  intensity-limited techniques that demand reducing the neutron losses
  between source and detectors. In addition, the nuclear industry
  demands more accurate data and procedures for the design and
  optimization of advanced fission reactors, especially for the
  treatment of fuel and moderator materials. To meet these demands, it
  is necessary to improve the existing calculation tools, through the
  generation of better models that describe the interaction of
  neutrons with the systems of interest. The Neutron Physics
  Department at Centro Atomico Bariloche (CNEA, Argentina) has been
  developing over the time new models for the interaction of slow
  neutrons with materials, to produce scattering kernels and cross
  section data in the thermal and cold neutron energy region. Besides
  the studies carried out on neutron moderators, we have recently
  begun looking at materials that could serve as efficient neutron
  reflectors over those energy ranges. In this work we present the
  results of transmission and scattering experiments on diamond
  nanopowder and magnesium hydride, carried out simultaneously at the
  VESUVIO spectrometer (ISIS, UK), and compare them with newly
  generated cross-section libraries.
\end{abstract}

\begin{keyword}
\kwd{Nano-diamond}
\kwd{Neutron reflection material} 
\kwd{Below cold neutron energy}
\kwd{Total neutron cross section}
\end{keyword}

\end{frontmatter}


\section{Introduction}\label{s1}
The tremendous potential of neutrons as a probe of matter or as a
research object by itself is limited by the relatively low-intensity
flux of neutron sources, as compared with photon sources.  In
addition, the different processes (production, slowing-down,
thermalization) and devices (moderators, transport systems,
collimators, energy selectors, detectors, etc.) reduce by several
orders of magnitude the actual neutron intensity that eventually
conveys the experimental information of interest.  For
  this reason, all optimization tasks that can be carried out on the
  devices involved are particularly relevant. This, in turn, motivates
  the creation of new and more precise models of neutron intraction
  with the systems of interest. The Neutron Physics Department at
Centro Atomico Bariloche (CNEA, Argentina) has been developing new
models for the interaction of slow neutrons with materials,
particularly those of interest for thermal and cold neutron sources.
Our aim is to produce scattering kernels and cross section data for
the corresponding energy range.  The approach involved the
determination of the excitation frequency spectra for liquid and solid
materials, employing Molecular Dynamics and ab initio calculations, in
combination with processing codes (NJOY, NCrystal).  Recently we
initiated a research line oriented to the search for efficient
reflector materials, that may improve the efficiency of guiding
surfaces or the actual reflection of neutrons on a containment wall to
reduce leakage.  A large body of work has been done in the past on
that quest, particularly concerning the interaction of slow neutrons
with diamond nanoparticles \cite{LYCHAGIN2009}.  The high reflectivity
of this material for has been demostrated for ultra-cold
neutrons (UCN) and very-cold neutrons (VCN) based on calculations for
ideal systems as well as by scattering experiments, proposing that
such capacity may extend to higher neutron energies, thus bridging the
``reflectivity gap'' in the neutron spectrum \cite{NESVIZHEVSKY2018}.
In a recent work we presented a new scattering kernel based on the
combination of bulk diamond particles and coherent interference
effects due to the nanostructure \cite{Granada2020}.  With those
ingredients we generated a cross section library over an extended
neutron energy range for a real nanodiamond (ND) powder.  Reflectivity
calculations using our library confirm the excellent performance of
that material for UCN and VCN, as well as its rapid decrease for
neutron velocities larger than about 150 m/s.

The reduction of reflectivity in the nanopowder is of course related
to the disappearance of the interference contribution to the
scattering process as the neutron wavelength becomes shorter than the
particle's dimension.  Therefore, a different approach was employed to
explore the characteristics of potential cold neutron reflectors,
based on the requirement of a large scattering cross section combined
with a small value of the average cosine of the scattering angle over
the energy range of interest ($E_n \geq 10^{-4}$ eV).  Those
conditions led us to consider metal hydrides, and eventually select
magnesium hydride as a good candidate to fulfill those requirements
\cite{Granada2020}.  We have also recently produced a refined
scattering kernel for H in MgH$_2$, \cite{Granada2020} and compared
the calculated total cross sections with available experimental data
at room temperature, obtaining an encouraging agreement.  We
considered it necessary at this stage to perform specific measurements
to experimentally validate the predictions previously stated.  In this
respect VESUVIO is an almost ideal instrument to simultaneously
determine the transmission and scattering properties of diamond
nanoparticles and magnesium hydride.  In this work we present the
results of experiments on diamond nanopowder and magnesium hydride,
and compare them with the newly generated cross-section libraries.

\section{Scattering kernels}\label{s2}

\subsection{Scattering kernel for diamond nanoparticles}\label{s2.1}
Studies on diamond nanopowder have been carried out by Nesvizhevsky
and coworkers for quite a long time, involving calculations and
measurements on that material to support its use as an excellent
reflector for ultra- and very-cold neutrons \cite{Nesvizhevsky2002}
and good quasi-specular reflector for cold neutrons \cite{Nesvi2018}.
As the main interest in those studies was the understanding and
description of the dominant (elastic) interference effects caused by
the scattering particles, those calculations were done for ideal ND
systems using the first Born approximation \cite{Artemev2016}.
Recently, SANS and transmission measurements were performed
\cite{TESHIGAWARA2019}, producing the first experimental total cross
section results for a diamond nanopowder over the cold neutron energy
region.

For the development of our model we consider a system of nanospheres
made of bulk diamond.  The scattering kernel will involve
contributions due to the elastic and inelastic scattering by the
diamond lattice, and also the coherent elastic contribution arising
from the diffraction of the neutron wave caused by the scattering
nanoparticles.  The phonon density of states (PDOS) of diamond was
evaluated by Density Functional Theory techniques
(DFT)\cite{Giannozzi2017}, showing excellent agreement with the
experimental results when used to calculate the specific heat
\cite{Victor1962}.  We then employed the code NJOY
\cite{macfarlane2017njoy} for the inelastic cross sections and
NCRYSTAL \cite{cai2020ncrystal} for the elastic ones.  

\begin{figure}[t]
\resizebox{0.6\textwidth}{!}{\includegraphics{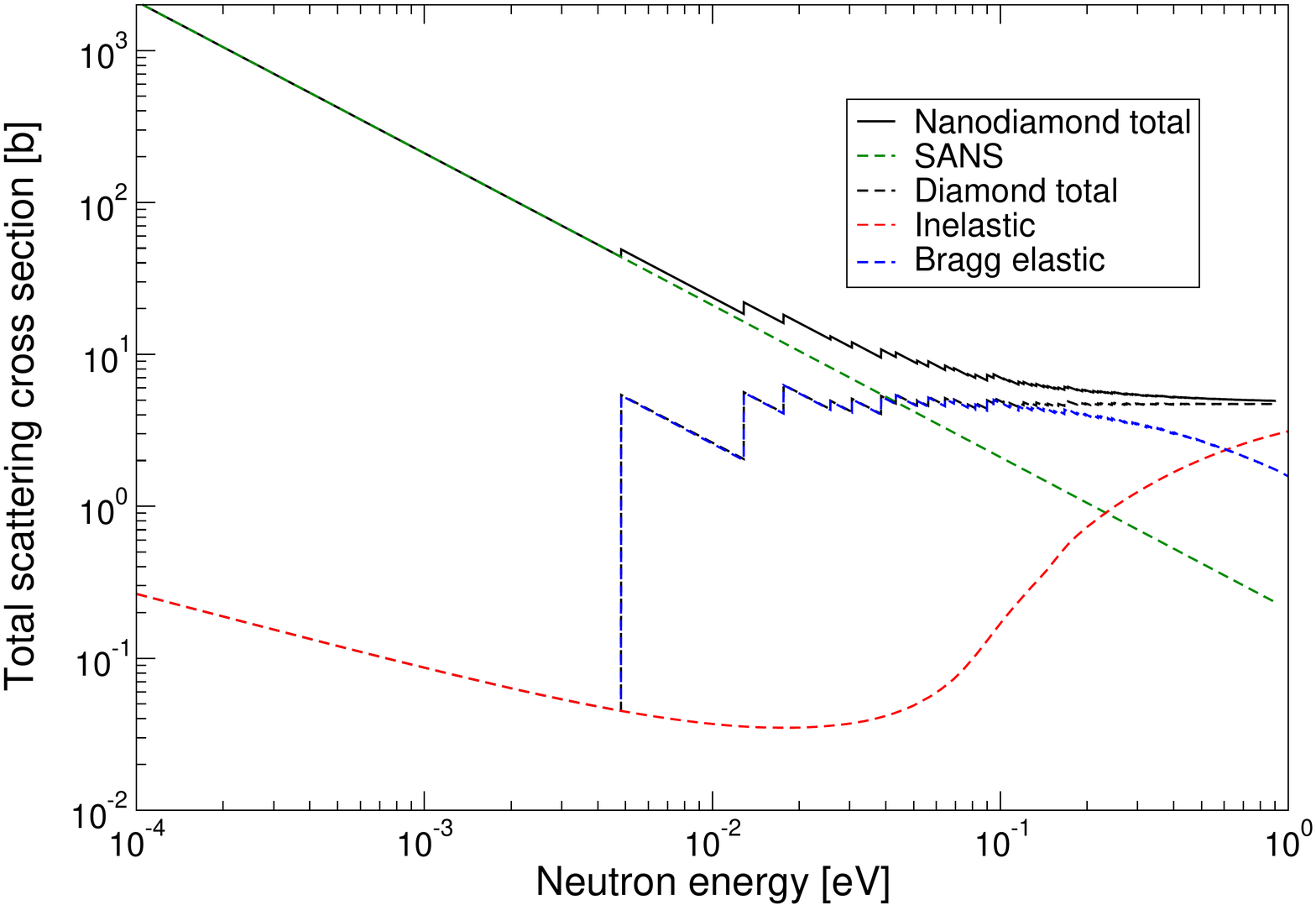}}
\caption{Total scattering cross section for diamond nanoparticle
  powder. See text for details. \label{ND_model}}
\end{figure}

We used a structure factor expression built from the unified
exponential/power-law approximation \cite{Beaucage} to represent the
interference of the neutron-wave caused by the finite size of the
(spherical) scattering potential.  The structure factor can be
integrated to obtain the total scattering cross section due to such
interference effects, which presents a $1/E$ behavior for neutron
energies above $\approx 2.10^{-5}$ eV. Assuming perfect spherical
diamond particles, a simple expression can be derived from the results
of the first Born approximation \cite{Artemev2016}:
\begin{equation}
\label{eqsans}
    \sigma(E)=0.006872\frac{R^{4.015}}{E},
\end{equation}
where $R$ is the average diamond sphere's radius in nm, $E$ the
neutron energy in eV, and $\sigma$ the total coherent cross section in
barns. The above formula is accurate enough to determine
a reliable value of $R$ from a total cross section measurement. The
calculated total cross section of nanodiamond powder at room
temperature with $R$=2.5 nm is shown in Fig. \ref{ND_model}, where the
different contributions of the crystal and SANS are indicated.

\subsection{Scattering kernel for magnesium hydride}
\label{s2.2}

We have proposed recently \cite{Granada2020} that MgH$_2$ could
be a very interesting material as reflector for cold neutrons, taking
advantage of some unique features of the hydrogen in this system:
large scattering cross section, high number density, practically no
translational motion within the metallic lattice, and unpopulated
(acoustic and optical) modes at low temperatures.  The latter
dynamical characteristics imply a very large effective mass and
therefore almost isotropic scattering, and a very small upscattering
probability, respectively. We produced a refined scattering kernel for
H in MgH$_2$, \cite{Granada2020} and compared the calculated total
cross sections with available experimental data at room temperature,
obtaining very good agreement.
\begin{figure}
    \centering
    \resizebox{0.6\textwidth}{!}{\includegraphics{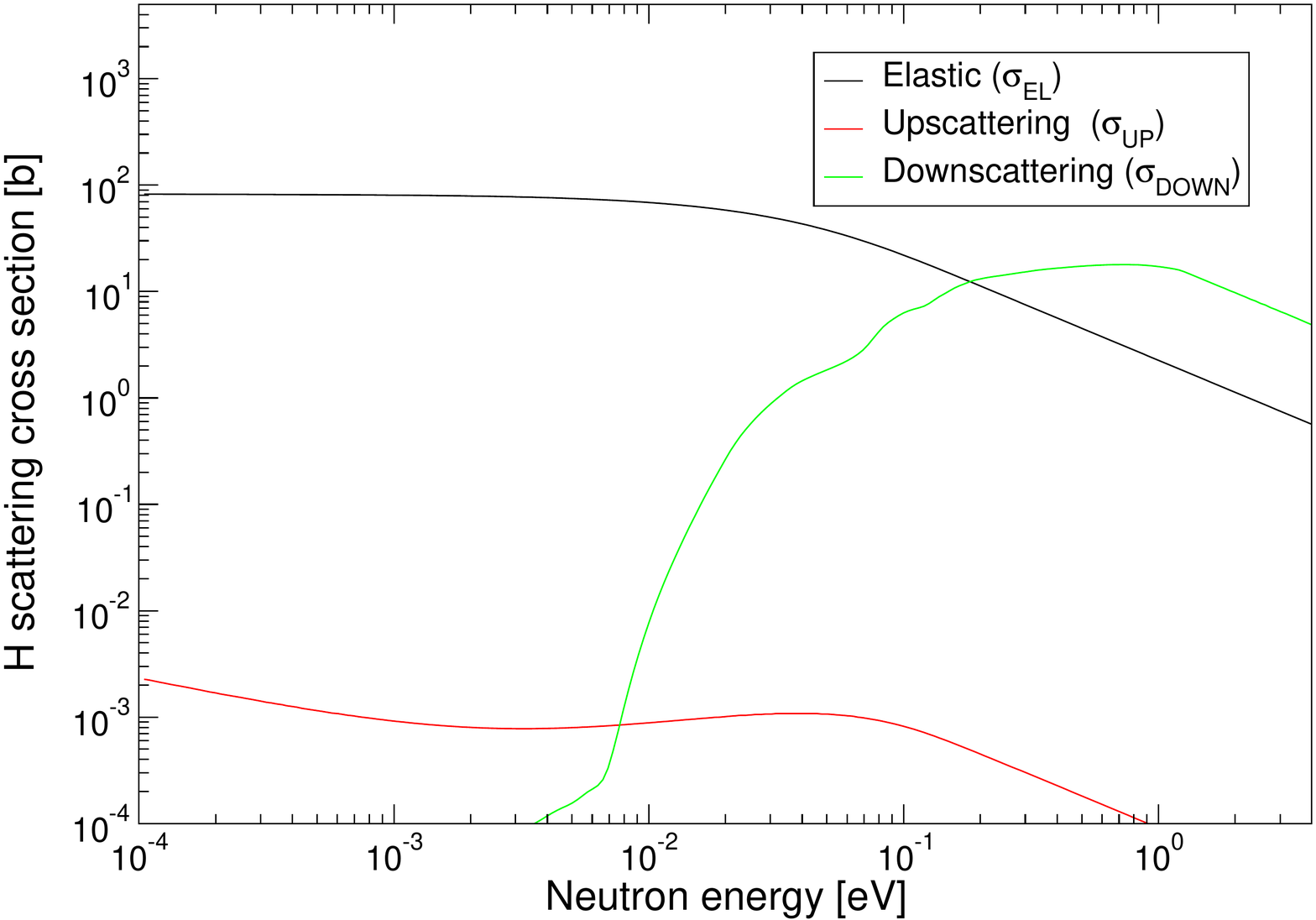}}
    \caption{Comparison of the scattering components for MgH$_2$ at
      20K. $\sigma_\text{EL}$ is the elastic component, and the inelastic
      part is separated into up-scatter($\sigma_\text{UP}$) and
      down-scatter($\sigma_\text{DOWN}$). }
    \label{fig:mgh2}
\end{figure}
The elastic and inelastic (up and down) scattering components of the
total cross section of H in MgH$_2$ at 20K are displayed in Fig.
\ref{fig:mgh2}.  Note that elastic processes dominate the interaction
for cold neutrons.  This behaviour is due to the absence of low-energy
excitation modes that causes the inelastic down-scattering processes
to be reduced rapidly below thermal neutron energies.

\section{Experimental details}\label{s3}

The experiments were performed at the VESUVIO spectrometer (ISIS
Pulsed Neutron and Muon Source at the Rutherford Appleton Laboratory,
United Kingdom).  The instument is placed at Target Station TS1.  This
is a versatile instrument that was originally designed for Deep
Inelastic Neutron Scattering and has proven to be very suitable for
the measurement of transmissions \cite{cal}.  The sample-source
flight-path is about 11 m. The detection system consists of scattering
and transmission detectors.  The former are placed in forward and
backward scattering directions.  In the forward-scattering directions
the detector banks are alternated in two characteristic flight-path
lengths of about 0.53 and 0.72 m from the sample position, covering an
angular range from 30$^{\circ}$ to 70$^{\circ}$, whereas in the
backward scattering directions the detectors are arranged in three
plane panels with sample-detector distances ranging from 0.44 m (for
those placed at the inner positions) to 0.65 m (for those placed at
the peripheral positions), with angles from 131$^{\circ}$ to
166$^{\circ}$.  The forward scattering detectors are labelled S135 to
S198, and the backscattering ones S3 to S134. A schematic
representation of the instrument and additional pieces of information
are available in Ref.~\cite{2017_Romanelli_MST}.

For the transmission measurements a detector labeled S2 is placed
at 13.43~m from the neutron moderator.  A beam monitor S1 is exposed
at the direct beam, at 8.60~m from the moderator, and its count rate
is employed to normalise the spectra registered by the transmission
monitor. Both detectors are $^6$Li-doped glass scintillators.

The samples employed in this work were Carbon Nanodiamond and
Magnesium Hydride powders. Nanodiamond powder was manufactured by Ray
Techniques Ltd. (Israel). It was produced from detonation synthesis
and purified. It presents the aspect of a gray powder with a nominal bulk
density of 2.7 g/cm$^3$ with an average crystalline size of 4.3
nm. Metal impurities were not detected and chlorine impurities were
less than 1 wt.\%.  The magnesium hydride sample under study was a white micro powder 99\%
purity produced by Nanoshel company (USA). The average particle size
was 40 to 60 $\mu$m, and its bulk nominal density 1.45 g/cm$^3$.

The samples were placed into aluminum flat cells of square section
64~mm inner side and 90~mm outer side. For Magnesium Hydride a 4 mm
thick cell was chosen, while for Carbon Nano Diamonds thicknesses of
4mm and 7 mm were used. Samples were placed into a standard closed
cycle refrigerator (CCR), which allowed to control the temperature
during measurements.

\section {Results}
\label{results}
\subsection{Data processing and results for diamond
  nanoparticles}\label{s3.2}

Transmitted spectra with and without the sample were processed
according to \cite{robledo2020measurement}, using the S1 monitor for
normalization.  The data processed directly from the transmission
measurements using the Beer-Lambert law show a significant departure
from the theoretical model, which suggests that a non-negligible
number of the scattering events are counted due to small angle
scattering.

\begin{figure}[H]
\resizebox{0.6\textwidth}{!}{\includegraphics{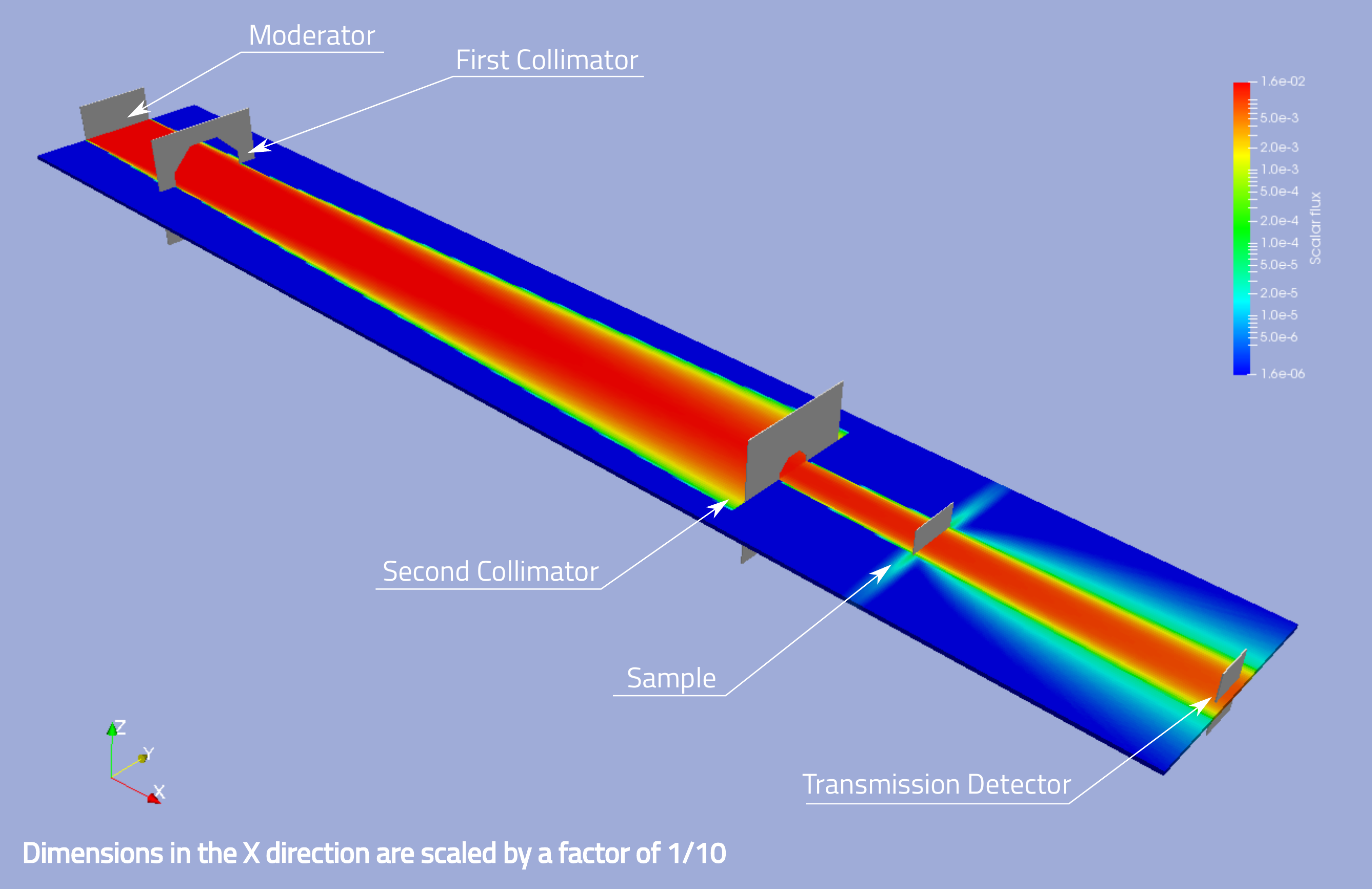}}
\caption{Flux map computed with the Monte Carlo model. The intensity
  is normalized to one source particle.}\label{MC}
\end{figure}

To take this effect into account, we performed a simulation of the
instrument in Monte Carlo. Geometry was adapted from
\cite{di2018mcstas}, and consists on a slab source in the position of
the moderator ($1101$ cm from the sample position), and two octogonal
collimators at $930$ and $135$ cm from the sample position, resulting
in a circular beam with $4.3$ cm diameter in the sample position. The
transmission detector was modelled as a $3.5 \times 5.6 \times 1$
cm$^3$ slab $244$ cm downstream from the sample. Fig. \ref{MC} shows
the flux distribution along the midplane of the instrument,
superimposed to the geometry. Scalar fluxes are normalized to one
source neutron.

Using this Monte Carlo model the contribution of small angle
scattering to the detector count was computed, and applied as a
correction for sample-dependent background:

\begin{equation}
  \label{inscatt}
\sigma(E) = -\dfrac{1}{N\Delta x}\log \dfrac{S-I}{E}
\end{equation}

\noindent where $N$ is the number density of atoms in the sample,
$\Delta x$ the sample thickness, $S$ and $E$ the normalized count for
sample and empty container, and $I$ the normalized count for
in-scattering events.

\begin{figure}[t]
\resizebox{0.6\textwidth}{!}{\includegraphics{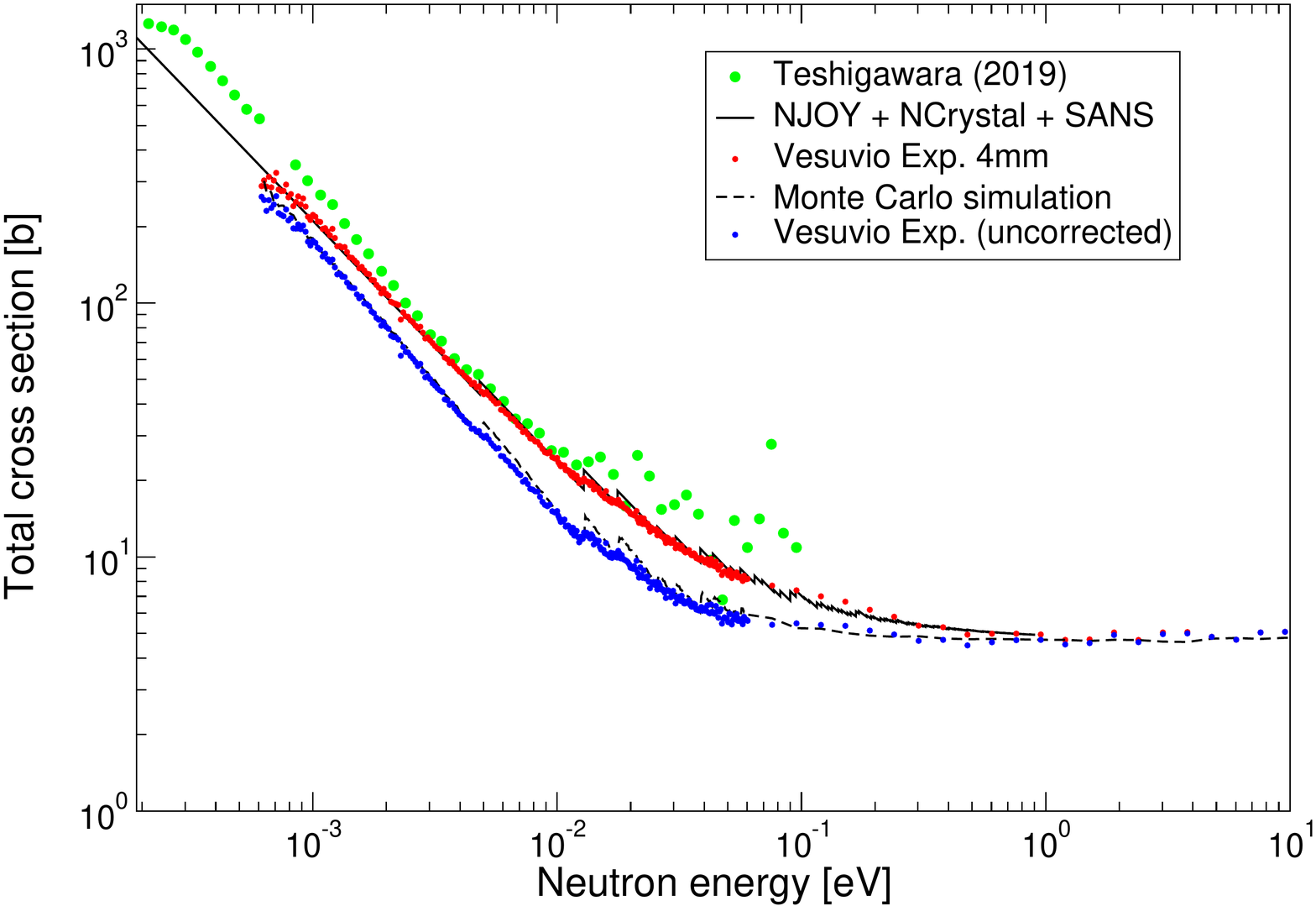}}
\caption{Experimental total scattering cross section for diamond
  nanoparticle powder. Red circles show the results computed after the
  in-scattering correction is applied, and the blue circles were
  computed with the Beer-Lambert law without taking into account
  in-scattering. Green circles are previous measurements by
  Teshigawara et. al
  \cite{TESHIGAWARA2019}.\label{ND_XS} In full line we
  show the calculated total cross section employing the code NJOY for
  the inelastic cross sections, NCRYSTAL for the elastic
  ones and the SANS component described in Eq. (\ref{eqsans}), while
  in dotted line the total cross section was calculated from the spectra
  determined by Monte Carlo simulations on the experimental setup.}
\end{figure}

\begin{figure}[t]
\resizebox{0.6\textwidth}{!}{\includegraphics{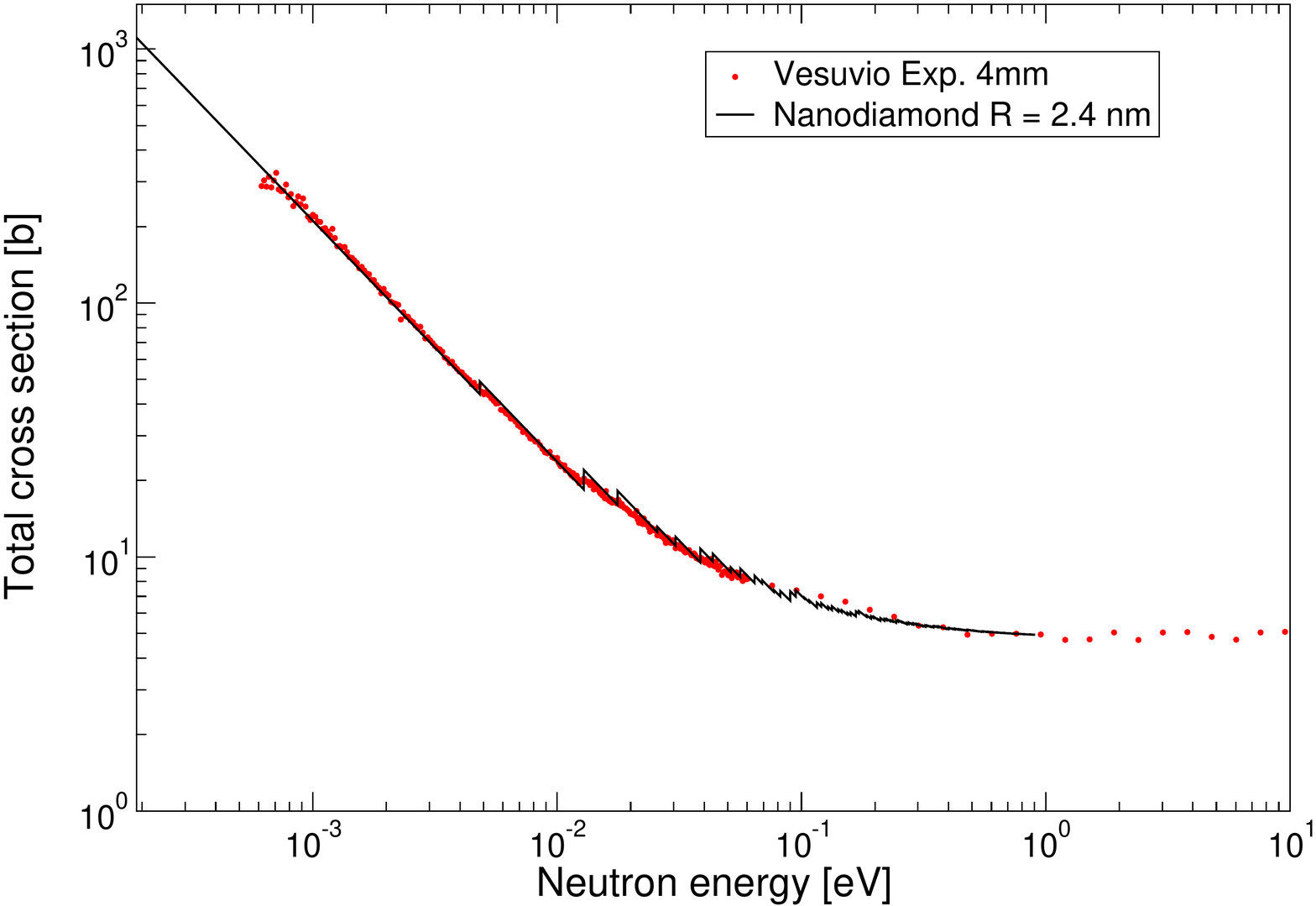}}
\caption{Corrected experimental cross section compared with our model
  calculations for an average nanodiamond sphere radius of 2.4 nm.}
\label{Comp_nd}
\end{figure}

 Fig. \ref{ND_XS} shows different experimental and theoretical results
compared. Blue circles represent the experimental total cross section
of the 4 mm thick ND sample computed with the Beer-Lambert law without
taking into account in-scattering. When this correction is
applied we obtain the data represented by the red circles.  A good
agreement is found between them and the above mentioned theoretical
model (full black line) elaborated by employing the NJOY code for the
inelastic cross sections, NCRYSTAL for the elastic ones and the SANS
component described in Eq. (\ref{eqsans}) computed with the value $R$
= 2.4 nm.  Also there is agreement with the experimental results of
Teshigawara et al \citep{Nesvizhevsky2002} (green cicles). In
Fig. \ref{ND_XS} we also show the total cross section calculated from
the spectra determined by Monte Carlo simulations on the experimental
setup (dashed black line), which are in agreement with the blue
circles.  To highlight the importance of the result,
Fig. \ref{Comp_nd} shows separately the corrected experimental total
cross section of our nanodiamond sample compared with our model
calculations.

\subsection{Data processing and results for magnesium hydride}\label{s3.3}

\begin{figure}[H]
\resizebox{0.6\textwidth}{!}{\includegraphics{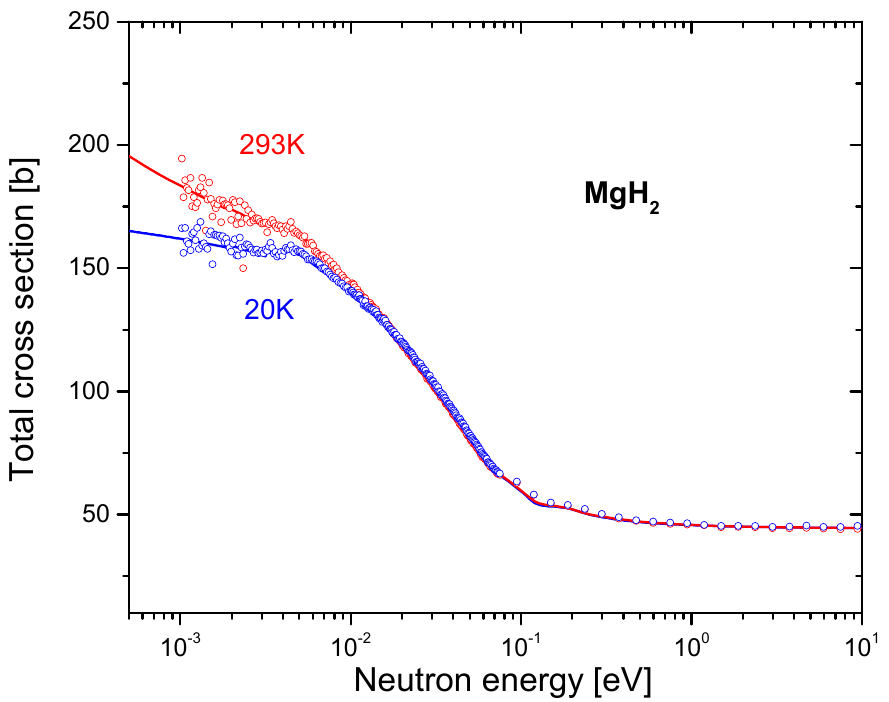}}
\caption{Experimental total scattering cross section for
  $MgH_2$ at two temperatures, 20K (blue) and 293K (red). Theoretical
  models are plotted in continuous lines with their respective
  colors.}\label{fig/MgH2_XS}
\end{figure}

Transmitted spectra with and without the different MgH$_2$
samples were also processed according to \cite{robledo2020measurement}
using the S1 monitor for normalization. Data were processed using the
Beer-Lambert equation and the total cross section was obtained as

\begin{equation}
    \sigma_T(E) = - \frac{1}{N\Delta x} \ln \left(\frac{I_t(E,x)}{I_0(E)}\right),
\end{equation}
where $I_t(E,x)$ is the transmitted intensity through a sample of
thickness $x$ and number density $n$ (number of atoms per unit volume
in the sample), and $I_0(E)$ is the incident intensity of neutrons
with energy $E$. Two MgH$_2$ samples were measured at different
temperatures, one at room temperature and another one at 20K. The
resulting total cross sections are displayed in
fig. \ref{fig/MgH2_XS}. As expected, the theoretical model is in good
agreement with the experimental results without any extra
corrections. This is in agreement with the usual
  situation in VESUVIO, where for samples without an important
  small-angle scattering cross section, the signal from scattered
  neutrons is negligible in the transmission
  detector~\cite{2019_Capelli_JAC}.

\subsection{Comparing reflectivities}
\begin{figure}[H]
\resizebox{0.6\textwidth}{!}{\includegraphics{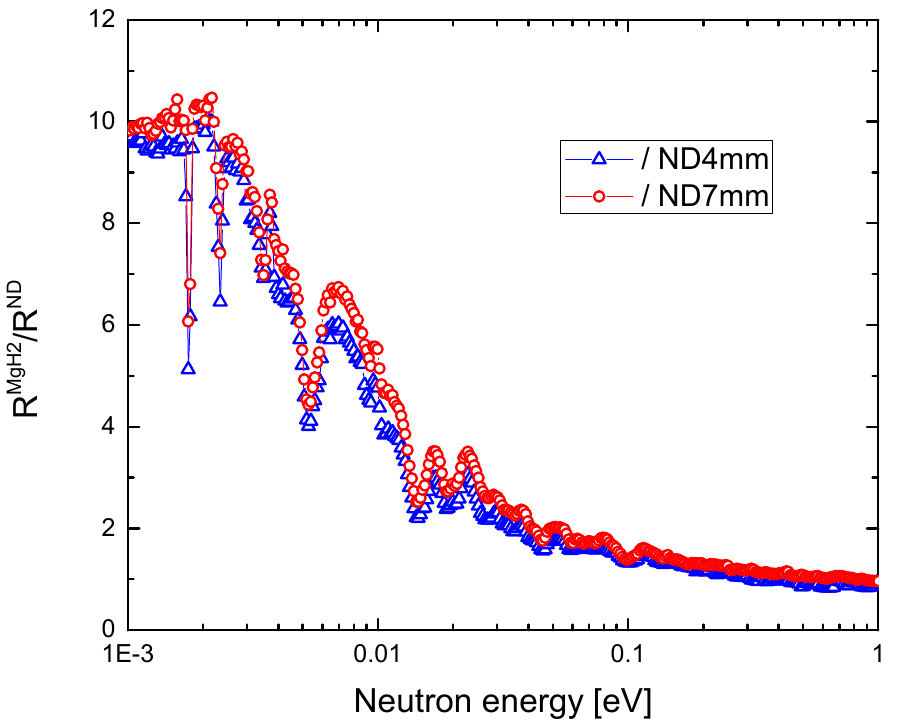}}
\caption{Energy distribution of the ratio of reflectivities between
  $MgH_2$ at 20 K and the nanodiamond sample of 4mm (blue)
  and 7mm (red) thickness.}\label{fig/reflect}
\end{figure}

The relationship stated between the number of neutron counts with
energy $E$ of a detector set at an angle $\theta$ for sample $i$
($C^i (\theta,E)$) and its reflectivity $R^i (\theta,E)$ is
\begin{equation}
C^i(\theta) = \phi(E) \left( 1 - Tr^i(E)\right) R^i (\theta,E) \varepsilon(E), 
\end{equation} 
where $\phi(E)$ is the incident flux, $\varepsilon(E)$ is the detector
efficiency, and $ \left( 1 - Tr^i(E)\right)$ the fraction of neutrons
not transmitted through the sample. In order to compare the reflective
power of both compounds, a ratio proportional to the ratio of their
reflectivities was calculated for each detector as
\begin{equation}\label{ec/cociente_C}
  \frac{R^{MgH_2}(\theta,E)}{R^{ND}(\theta,E)} = \frac{(1-Tr^{ND}(E))}{(1-Tr^{MgH_2}(E))}  \frac{C^{MgH_2}(\theta,E)}{C^{ND}(\theta,E)} ,
\end{equation} 
where $R^i (\theta,E)$ denotes a quantity proportional to the
reflectivity of the $i^{th}$ compound (ND refers to nanodiamond) at an
angle $\theta$ for neutrons with energy $E$, and $C^i (\theta,E)$ the
corresponding counts of the detector set at an angle $\theta$ with
respect to the incident beam. To obtain $C^i (\theta,E)$, the
background contribution was subtracted to each detector signal by measuring
an empty can sample, and a normalization was performed to the incident
beam intensity. The ratio was performed channel by channel using the
Mantid framework \cite{Arnold2014}. The resulting ratio of
reflectivities averaged through the detectors taking into account
equation (\ref{ec/cociente_C}) is shown in Fig. \ref{fig/reflect}.

\begin{figure}[H]
\resizebox{0.8\textwidth}{!}{\includegraphics{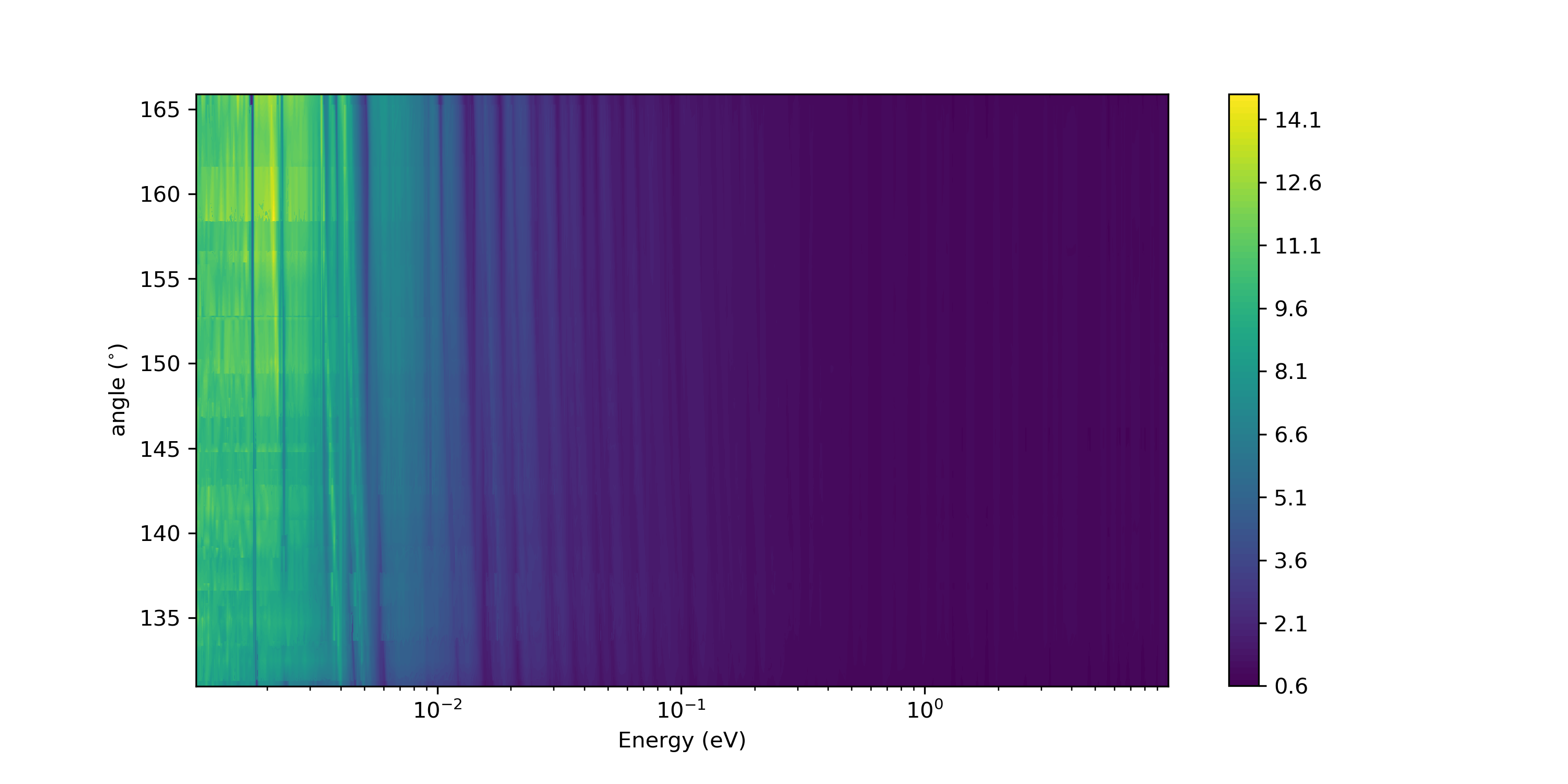}}
\caption{Energy and angular distribution of the ratio of
  reflectivities between MgH$_2$ at 20K and the
  nanodiamond sample of 7mm thickness for the back-scattering detector
  banks.}\label{fig/2dmap}
\end{figure}

The peak structure seen in Fig. \ref{fig/reflect} between 1 and 100
meV is due to the presence of sharp Bragg peaks in the nanodiamond
sample. Despite this structure, it is interesting to note that the
ratio of reflectivities as defined in this work is
$R^{MgH2}/R^{ND}\geq 1$ for all energy values below 1000 meV, implying
that $MgH_2$ is a more efficient reflector than nanodiamonds for
thermal and cold neutrons.  As to emphasize the angular dependency, a
2D map of the angle and energy distribution of the ratio of
reflectivities is shown in Fig. \ref{fig/2dmap}, where the
displacement of Bragg peaks with the measuring angle is apparent.

\begin{figure}[t]
\resizebox{0.6\textwidth}{!}{\includegraphics{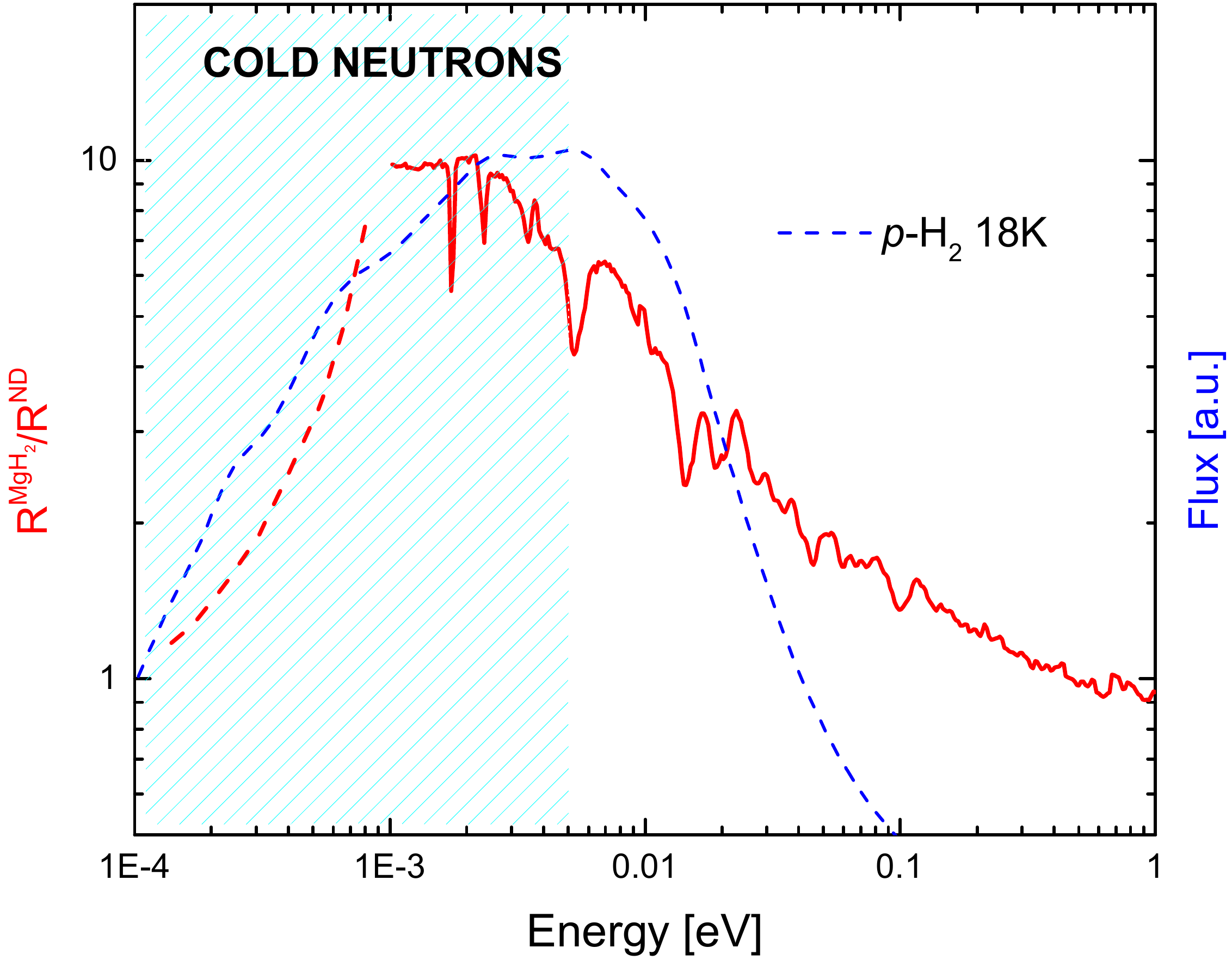}}
\caption{Typical cold neutron flux emerging from a liquid hydrogen
  source (blue dashed line) and the ratio of reflectivities between cold MgH$_2$ and ND (red line). The dashed red line represents the expected trend of the reflectivity ratio (Eq(\ref{ec/cociente_C})) in the cold neutron region.}
\label{Cold_flux}
\end{figure}

It is expected \cite{charles} that the reflectivities of both
materials are similar at about 10$^{-4}$ eV, and that below that energy
the ND will clearly be better. However, as in most current applications
we can consider that liquid H$_2$ is the cold neutron moderator and we
wish to reflect back into the moderator cell those neutrons leaking in
the wrong direction. In Fig. \ref{Cold_flux} we compare a typical cold
neutron flux emerging from a liquid hydrogen source and the ratio of
reflectivities between cold Magnesium Hydride and ND determined in this work, 
extrapolated with the expected trend of the reflectivity ratio in the cold neutron region. Magnesium Hydride
is much more efficient than the diamond nanopowder to reflect neutrons
produced by a cold source, although the latter should be an excellent
coating material to transport cold neutrons on account of its very large quasi
specular reflection capacity \cite{jamalipour}.

\section{Conclusions}\label{s4}

We performed transmission and scattering experiments at room and low
temperatures on nanodiamond and MgH$_2$ powder samples, using the
VESUVIO instrument at the ISIS pulsed neutron source.

The total cross sections for neutron energies 10$^{-3}$ to 10 eV were
determined from the transmission measurements on both
materials. Especially for the ND samples, a careful in-scattering
correction procedure was required, on account of the very large SANS
contribution caused by interference effects. We calculated the total
cross sections using our scattering kernels and the cross section
libraries generated from them, showing an excellent agreement with the
corrected experimental total cross sections.

The scattering measurements were used to make relative comparison of
the reflecting capacity of ND and MgH$_2$. The different sets of
detectors were grouped to facilitate the comparison with good
statistics. An expression proposed to consider the ratio of
reflectivities shows to be adequate to describe the material
properties irrespective of the sample thickness employed in the
comparison. The results confirm the better performance of MgH$_2$ at
low temperatures as compared with nanodiamonds over the ``cold neutron
reflectivity gap'', as we have previously predicted from theoretical
considerations.

The actual gain obtained by the use of such reflector will depend on the specific application, and should become more noticeable in systems with high leakage, as in a low dimensional liquid hydrogen source.

\section{Acknowledgements}

The authors acknowledge the UK Science and Facilities Council for the
beam time assigned at the ISIS Neutron Source - Rutherford Appleton
Laboratory (VESUVIO, https://doi.org/10.5286/ISIS.E.RB1920136).





\nocite{*} 

\bibliographystyle{ios1}           
\bibliography{bibliography}        

%

\end{document}